\documentclass[aps,preprint]{revtex4}%
\usepackage{amsfonts}
\usepackage{amsmath}
\usepackage{amssymb}
\usepackage{graphicx}%
\providecommand{\U}[1]{\protect\rule{.1in}{.1in}}

\begin{document}
\title{{\LARGE RELATIVISTIC COSMOLOGY AND \ THE PIONEERS ANOMALY}}
\author{Marcelo Samuel Berman$^{1}$ and \ Fernando de Mello Gomide$^{1}$}
\affiliation{$^{1}$Instituto Albert Einstein/Latinamerica\ - Av. Candido Hartmann, 575 -
\ \# 17}
\affiliation{80730-440 - Curitiba - PR - Brazil }
\affiliation{emails: msberman@institutoalberteinstein.org }
\affiliation{\ \ \ \ \ \ \ \ \ \ marsambe@yahoo.com}
\date{22 July 2011}

\begin{abstract}
NASA spacecrafts has suffered from three anomalies. The Pioneers spacecrafts
were decelerated, and their spin when not disturbed, was declining. On the
other hand, fly-bys for gravity assists, appeared with extra speeds, relative
to infinity. The Pioneers and fly-by anomalies are given now exact general
relativistic full general solutions, in a rotating expanding Universe.We cite
new evidence on the rotation of the Universe. Our solution seems to be the
only one that solves the three anomalies.

\end{abstract}
\maketitle

{\LARGE RELATIVISTIC COSMOLOGY AND \ THE PIONEERS ANOMALY }

\begin{center}
\bigskip Marcelo Samuel Berman and Fernando de Mello Gomide

\end{center}

\bigskip{\LARGE 1. Introduction\qquad\ }

Anderson et al.(2008), and L\"{a}mmerzahl et al.,2006), have alerted the
scientific community about the fly-by anomaly: during Earth gravity assists,
spacecraft has suffered from an extra -energization, characterized by a
positive extra speed, such that, measured "at infinity", the hyperbolic
orbiting \ object presented an empirically calculated \ \ $\Delta V/V$
\ \ \ around \ $10^{-6}.$A formula was supplied,

$\frac{\Delta V}{V}=\frac{2\omega R}{c}$ \ \ \ \ \ \ \ \ \ \ \ \ \ \ \ \ \ \ \ \ \ \ \ \ \ \ \ \ \ \ \ \ \ \ \ \ \ \ \ \ \ \ \ \ \ \ \ \ \ \ \ \ \ \ \ (1.1)

\bigskip where \ \ $\omega$ \ \ , $\ \ \ R$ \ \ \ and \ \ $c$ \ \ \ \ \ stand
for the \ angular speed and radius of the central mass, and the speed of light
in vacuo. T.L. Wilson, from NASA,(Houston) and H.-J. Blome (Aachen), delivered
a lecture in Montreal , on July 17,2008, and called the attention to the fact
that the most trusted cause for \ both this anomaly, and the Pioneers, would
be "rotational dynamics"(Wilson and Blome ,(2008). One of us, had, by that
time, published \ results on the Pioneers Anomaly, through the rotation of the
Universe (Berman, 2007). Now, we shall address the three anomalies.

The Pioneers Anomaly is the deceleration of \ about -$9.10^{-8}$cm.s$^{-2}$
suffered by NASA space-probes travelling towards outer space. It has no
acceptable explanation within local Physics, but if we resort to Cosmology, it
could be explained by the rotation of the Universe. Be cautious, because there
is no center or axis of rotation. We are speaking either of a Machian or a
General Relativistic cosmological vorticity. It could apply to each observed
point in the Universe, observed by any observer. Another explanation, would be
that our Universe obeys a variable speed of light Relativistic Cosmology,
without vorticities.However, we shall see later that both models are
equivalent. Thermal emission cannot be \ invoked, for it should also
decelerate elliptical orbiters, but the deceleration only affects hyperbolic
motion. It does not explain fly-bys, either. A secondary Pioneers anomaly
refers to spinning down of the spacecraft, when they were not disturbed.
Again, thermal emission cannot explain it.

\bigskip In previous papers (Berman and Gomide, 2010; 2011; 2011a), by
considering an exact but particular solution of Einstein%
\'{}%
s field equations for an expanding and rotating metric,found, by estimating
the deceleration parameter of the present Universe, as \ \ \ $q\approx-1/2$
\ \ \ \ , that the Universe appeared \ to possess a field of decelerations
coinciding approximately with the Pioneers anomalous value (Anderson et al.,
2002).We now shall consider the condition \ for an exact match with the
Pioneers deceleration,with a large class of solutions in General Relativity.
Sections 5 and \ 6, treat the second Pioneers anomaly, and the fly-by. In
section 7, an alternative cosmological model will be presented, following an
idea by Godlowski et al.(2004), which allows us to work with a non-modified RW%
\'{}%
s metric.

\bigskip The key result for all these subjects, is that hyperbolic motion,
extends towards infinity, and, thus, qualify for cosmological alternatives,
and boundary conditions. The fly-bys, and the Pioneers, are in hyperbolic
trajectories, when the anomalies appear, so that Cosmology needs to be invoked.

Ni (2008;2009), has reported observations on a possible rotation of the
polarization of the cosmic background radiation, around 0.1 radians. As such
radiation was originated at the inception of the Universe, we tried to
estimate a possible angular speed or vorticity, by dividing 0.1 radians by the
age of the Universe , obtaining about 10$^{-19}$rad.s$^{-1}$. Compatible
results were obtained by Chechin(2010) and Su and Chu (2009).

The numerical result is very close to the theoretical estimate, by Berman (2007),

\bigskip

$\omega=c/R=3.10^{-18}$rad.s$^{-1}.$ \ \ \ \ \ \ \ \ \ \ \ \ \ \ \ \ \ \ \ \ \ \ \ \ \ \ \ \ \ \ \ \ \ \ \ \ \ \ \ \ \ \ \ \ \ \ \ \ \ \ \ \ \ \ \ \ \ \ \ \ \ \ \ \ \ (1.2)

\bigskip

where $\ \ \ c$ \ \ , $\ \ \ R$ $\ \ \ $represent the speed of light in
vacuum, and the radius of the causally related Universe.

\bigskip

\bigskip When one introduces a metric temporal coefficient \ \ \ $g_{00}%
$\ \ which is not constant, the new metric includes rotational effects. The
metric has a rotation of the tri-space (identical with RW%
\'{}%
s tri-space) around the orthogonal time axis. This will be our framework,
except for Section 7.

\bigskip

.

{\LARGE 2. On the four kinds of rotation in Relativistic Cosmology}

\bigskip

Consider the metric line-element:

\bigskip

$ds^{2}=g_{\mu\nu}dx^{\mu}dx^{\nu}$ \ \ \ \ \ \ \ \ \ \ . \ \ \ \ \ \ \ \ \ \ \ \ \ \ \ \ \ \ \ \ \ \ \ \ \ \ \ \ \ \ \ \ \ \ \ \ \ \ \ \ \ \ \ \ \ \ \ \ \ \ \ \ \ \ \ \ \ \ \ \ \ \ \ \ \ \ \ \ \ \ \ \ (2.1)

\bigskip

If the observer is at rest,

\bigskip

$dx^{i}=0$ \ \ \ \ \ \ \ \ \ \ \ \ \ ( \ $i=1,2,3$\ \ ) \ \ \ \ ,

\bigskip

while,

\bigskip

$dx^{0}=dt$ \ \ \ \ \ \ \ \ \ \ \ . \ \ \ \ \ \ \ \ \ \ \ \ \ \ \ \ \ \ \ \ \ \ \ \ \ \ \ \ \ \ \ \ \ \ \ \ \ \ \ \ \ \ \ \ \ \ \ \ \ \ \ \ \ \ \ \ \ \ \ \ \ \ \ \ \ \ \ \ \ \ \ \ \ \ \ \ \ \ \ \ \ \ \ \ (2.2)

\bigskip

This last equality defines a proper time; we called cosmic time, in Cosmology.

\bigskip

From the geodesics' equations, we shall have:

\bigskip

$\frac{d^{2}x^{i}}{ds^{2}}+\Gamma_{\alpha\beta}^{i}\frac{dx^{\alpha}}{ds}%
\frac{dx^{\beta}}{ds}=\Gamma_{00}^{i}$ \ \ \ \ \ \ \ \ \ \ \ . \ \ \ \ \ \ \ \ \ \ \ \ \ \ \ \ \ \ \ \ \ \ \ \ \ \ \ \ \ \ \ \ \ \ \ \ \ \ \ \ \ \ \ \ \ \ \ \ \ \ \ \ \ \ \ \ \ (2.3)

\bigskip

We then find:

\bigskip

$g^{ij}\frac{\partial g_{i0}}{\partial t}=0$ \ \ \ \ \ \ \ \ \ \ \ . \ \ \ \ \ \ \ \ \ \ \ \ \ \ \ \ \ \ \ \ \ \ \ \ \ \ \ \ \ \ \ \ \ \ \ \ \ \ \ \ \ \ \ \ \ \ \ \ \ \ \ \ \ \ \ \ \ \ \ \ \ \ \ \ \ \ \ \ \ \ \ \ \ \ \ \ \ \ \ \ (2.4)

\bigskip

This defines a Gaussian coordinate system, which in general implies that:

\bigskip

$\frac{\partial g_{i0}}{\partial t}=0$ \ \ \ \ \ \ \ \ \ \ \ . \ \ \ \ \ \ \ \ \ \ \ \ \ \ \ \ \ \ \ \ \ \ \ \ \ \ \ \ \ \ \ \ \ \ \ \ \ \ \ \ \ \ \ \ \ \ \ \ \ \ \ \ \ \ \ \ \ \ \ \ \ \ \ \ \ \ \ \ \ \ \ \ \ \ \ \ \ \ \ \ \ \ \ \ (2.5)

\bigskip

We must now reset our clocks, so that, the above condition is universal (valid
for all the particles in the Universe), and then our metric will assume the form:

\bigskip

$ds^{2}=dt^{2}-g_{ij}\left(  \vec{x},t\right)  dx^{i}dx^{j}$
\ \ \ \ \ \ \ \ \ \ \ . \ \ \ \ \ \ \ \ \ \ \ \ \ \ \ \ \ \ \ \ \ \ \ \ \ \ \ \ \ \ \ \ \ \ \ \ \ \ \ \ \ \ \ \ \ \ \ \ \ \ \ \ \ \ \ \ (2.6)

\bigskip

If we further impose that, in the origin of time, we have:

\bigskip

$g_{i0}(t=0)=0$ \ \ \ \ \ \ \ , \ \ \ \ \ \ \ \ \ \ \ \ \ \ \ \ \ \ \ \ \ \ \ \ \ \ \ \ \ \ \ \ \ \ \ \ \ \ \ \ \ \ \ \ \ \ \ \ \ \ \ \ \ \ \ \ \ \ \ \ \ \ \ \ \ \ \ \ \ \ \ \ \ \ \ \ \ (2.7)

\bigskip

then by (2.5), we shall have:

\bigskip

$g_{i0}(t)=0$ \ \ \ \ \ \ \ \ \ \ . \ \ \ \ \ \ \ \ \ \ \ \ \ \ \ \ \ \ \ \ \ \ \ \ \ \ \ \ \ \ \ \ \ \ \ \ \ \ \ \ \ \ \ \ \ \ \ \ \ \ \ \ \ \ \ \ \ \ \ \ \ \ \ \ \ \ \ \ \ \ \ \ \ \ \ \ \ \ \ \ \ (2.8)

\bigskip

The above defines a Gaussian normal coordinate system.

\bigskip

For a commoving observer, in a freely falling perfect fluid, the
quadrivelocity \ \ $u^{\mu}$\ \ will obey:

\bigskip

$u^{i}=0$ \ \ \ \ \ \ \ \ , \ \ \ \ \ \ \ \ \ \ \ \ \ \ \ \ \ \ \ \ \ \ \ \ \ \ \ \ \ \ \ \ \ \ \ \ \ \ \ \ \ \ \ \ \ \ \ \ \ \ \ \ \ \ \ \ \ \ \ \ \ \ \ \ \ \ \ \ \ \ \ \ \ \ \ \ \ \ \ \ \ \ \ \ \ \ \ \ \ \ \ (2.9)\ 

\bigskip

while, if we normalize the quadrivelocity, we find, from the condition:

\bigskip

$g_{\mu\nu}u^{\mu}u^{\nu}=1$ \ \ \ \ \ \ \ \ \ \ , \ \ \ \ \ \ \ \ \ \ \ \ \ \ \ \ \ \ \ \ \ \ \ \ \ \ \ \ \ \ \ \ \ \ \ \ \ \ \ \ \ \ \ \ \ \ \ \ \ \ \ \ \ \ \ \ \ \ \ \ \ \ \ \ \ \ \ \ \ \ \ \ \ \ \ \ \ \ \ (2.10)

\bigskip

that,

\bigskip

$g_{00}u^{0}=1$ \ \ \ \ \ \ \ \ \ . \ \ \ \ \ \ \ \ \ \ \ \ \ \ \ \ \ \ \ \ \ \ \ \ \ \ \ \ \ \ \ \ \ \ \ \ \ \ \ \ \ \ \ \ \ \ \ \ \ \ \ \ \ \ \ \ \ \ \ \ \ \ \ \ \ \ \ \ \ \ \ \ \ \ \ \ \ \ \ \ \ \ \ \ (2.11)\ 

\bigskip

Though later we shall discuss  the case \ \ \ $g_{00}=g_{00}(t)\neq1$\ \ \ ,
it is usually imposed:

\bigskip

$g_{00}=u^{0}=1$ \ \ \ \ \ \ \ \ \ \ \ . \ \ \ \ \ \ \ \ \ \ \ \ \ \ \ \ \ \ \ \ \ \ \ \ \ \ \ \ \ \ \ \ \ \ \ \ \ \ \ \ \ \ \ \ \ \ \ \ \ \ \ \ \ \ \ \ \ \ \ \ \ \ \ \ \ \ \ \ \ \ \ \ \ \ \ \ \ (2.12)\ \ 

\bigskip

When dealing with Robertson-Walker's metric, this is the usual procedure. By
this means, we have a tri-space, orthogonal to the time axis.

\bigskip

Gaussian coordinate systems, in fact, imply that, with \ \ $g_{0i}=0$\ \ ,
there are no rotations in the metric, and in each point we may define a
locally inertial reference system.

\bigskip

Gaussian normal coordinates were called "synchronous"\ ; in an arbitrary
spacetime, when we pick a spacelike\ \bigskip hypersurface \ \ $S_{0}$\ , and
we eject geodesic lines orthogonal to it, with constant coordinates\ \ $x^{1}%
,x^{2}$\ and \ \ $x^{3}$\ , while \ \ $x^{0}\equiv t+t_{0}$\ , where
\ $t_{0}=0$\ \ on \ $S_{0}$\ , then \ \ $t$\ \ is the proper time, whose
origin is \ \ $t=0$\ \ \ on \ \ $S_{0}$\ \ (see MTW, 1973)\ .\ \ \ \ 

\bigskip

In the above treatment, cosmic time is "absolute", so that the measure of the
age of the Universe, according to this "time", is not subject to a relative nature.

\bigskip

Now, we might ask whether the tri-space, orthogonal to the time axis, could
rotate relative to this axis. Berman (2008a, 2008b), has exactly defined this
original idea, by identifying this rotation, which is different from all
others, as will shall show bellow, with a time-varying metric coefficient
\ \ $g_{00}(t)$\ \ \ \ . In the next Section, we relate the angular speed of
the tri-space, relative to the time axis, with \ \ $g_{00}(t)$\ \ by means of,

\bigskip

$\omega=\frac{1}{2}\frac{\dot{g}_{00}}{g_{00}}$
\ \ \ \ \ \ \ \ \ \ \ \ \ \ \ \ \ \ \ \ \ \ \ \ \ \ \ \ \ \ \ \ \ \ \ . \ \ \ \ \ \ \ \ \ \ \ \ \ \ \ \ \ \ \ \ \ \ \ \ \ \ \ \ \ \ \ \ \ \ \ \ \ \ \ \ \ \ \ \ \ \ \ \ \ \ \ \ \ \ \ \ \ \ \ \ (2.13)

\bigskip In the above, we still may have a perfect fluid model.

\bigskip Other type of rotation, is Raychaudhuri's vorticity, which is
attached to non-perfect fluids (see, for instance, Berman (2007)). A third
type of rotation, is what we usually call rotation of the metric, and is
defined by non-diagonal terms, in the metric. For instance, Kerr's metric.

\bigskip

A fourth kind of rotation, is also attached to a perfect fluid model, like
Berman's one: it is the Godolowski et al. (2004) idea, which is developed in
Section 7 bellow.

.

{\LARGE 3. Field equations for Gomide-Uehara-R.W.-metric}

\bigskip

Consider first a temporal metric coefficient which depends only on \ $t$\ \ .
The line element becomes:\ 

\bigskip

$ds^{2}=-\frac{R^{2}(t)}{\left(  1+kr^{2}/4\right)  ^{2}}\left[  d\sigma
^{2}\right]  +g_{00}\left(  t\right)  $ $dt^{2}$
\ \ \ \ \ \ \ \ \ \ \ \ \ \ \ . \ \ \ \ \ \ \ \ \ \ \ \ \ \ \ \ \ \ \ \ \ \ \ \ \ \ \ \ \ \ \ \ \ \ \ \ \ \ \ \ \ \ \ \ \ \ (3.1)

\bigskip

The field equations, in \ General Relativity Theory (GRT) become:

\bigskip

$3\dot{R}^{2}=\kappa(\rho+\frac{\Lambda}{\kappa})g_{00}R^{2}-3kg_{00}$
\ \ \ \ \ \ \ \ \ \ \ \ \ \ \ \ \ \ \ \ \ , \ \ \ \ \ \ \ \ \ \ \ \ \ \ \ \ \ \ \ \ \ \ \ \ \ \ \ \ \ \ \ \ \ \ \ \ \ \ \ \ \ \ \ \ \ \ \ (3.2)

\bigskip

and,

\bigskip

$6\ddot{R}=-g_{00}\kappa\left(  \rho+3p-2\frac{\Lambda}{\kappa}\right)
R-3g_{00}\dot{R}$ $\dot{g}^{00}$ \ \ \ \ \ \ \ \ \ \ \ . \ \ \ \ \ \ \ \ \ \ \ \ \ \ \ \ \ \ \ \ \ \ \ \ \ \ \ \ \ \ \ \ \ \ \ \ \ \ \ \ \ \ (3.3)

\bigskip

Local inertial processes are observed through proper time, so that the
four-force is given by:

\bigskip

$F^{\alpha}=\frac{d}{d\tau}\left(  mu^{\alpha}\right)  =mg^{00}$ $\ddot
{x}^{\alpha}-\frac{1}{2}m$ $\dot{x}^{\alpha}\left[  \frac{\dot{g}_{00}}%
{g_{00}^{2}}\right]  $ \ \ \ \ \ \ \ \ \ \ \ \ \ \ . \ \ \ \ \ \ \ \ \ \ \ \ \ \ \ \ \ \ \ \ \ \ \ \ \ \ \ \ \ \ \ \ \ \ \ \ \ \ (3.4)

\bigskip

Of course, when \ $g_{00}=1$\ \ , the above equations reproduce conventional
Robertson-Walker's field equations.

\bigskip

In order to understand equation (3.4)\ , it is convenient to relate the
rest-mass $m$\ , \ \ to \ an inertial mass \ $M_{i}$\ , with:

\bigskip

$M_{i}=\frac{m}{g_{00}}$\ \ \ \ \ \ \ \ \ \ \ \ \ \ .\ \ \ \ \ \ \ \ \ \ \ \ \ \ \ \ \ \ \ \ \ \ \ \ \ \ \ \ \ \ \ \ \ \ \ \ \ \ \ \ \ \ \ \ \ \ \ \ \ \ \ \ \ \ \ \ \ \ \ \ \ \ \ \ \ \ \ \ \ \ \ \ \ \ \ \ \ \ \ \ \ \ \ \ \ \ \ (3.5)

\bigskip

It can be seen that \ $M_{i}$\ \ represents the inertia of a particle, when
observed along cosmic time, i.e., coordinate time. In this case, we observe
that we have two acceleration terms, which we call,

\bigskip

\bigskip\ $a_{1}^{\alpha}=\ddot{x}^{\alpha}$\ \ \ \ \ \ \ \ \ \ \ \ \ \ \ , \ \ \ \ \ \ \ \ \ \ \ \ \ \ \ \ \ \ \ \ \ \ \ \ \ \ \ \ \ \ \ \ \ \ \ \ \ \ \ \ \ \ \ \ \ \ \ \ \ \ \ \ \ \ \ \ \ \ \ \ \ \ \ \ \ \ \ \ \ \ \ \ \ \ \ \ \ \ \ \ \ \ \ \ \ \ \ \ (3.6)

\bigskip

and,

\bigskip

\bigskip\ $a_{2}^{\alpha}=-\frac{1}{2g_{00}}\left(  \dot{x}^{\alpha}\dot
{g}_{00}\right)  $%
\ \ \ \ \ \ \ \ \ \ \ \ \ \ \ .\ \ \ \ \ \ \ \ \ \ \ \ \ \ \ \ \ \ \ \ \ \ \ \ \ \ \ \ \ \ \ \ \ \ \ \ \ \ \ \ \ \ \ \ \ \ \ \ \ \ \ \ \ \ \ \ \ \ \ \ \ \ \ \ \ \ \ \ \ \ \ \ (3.7)\bigskip

\bigskip

The first acceleration is linear; the second, resembles rotational motion, and
depends on \ $g_{00}$\ \ and its time-derivative.\ \ 

\bigskip

If we consider \ \ $a_{2}^{\alpha}$\ \ a centripetal acceleration, we conclude
that the angular speed \ \ $\omega$\ \ \ is given by,

\bigskip

$\omega=\frac{1}{2}\left(  \frac{\dot{g}_{00}}{g_{00}}\right)  $%
\ \ \ \ \ \ \ \ \ \ \ \ \ \ \ \ \ . \ \ \ \ \ \ \ \ \ \ \ \ \ \ \ \ \ \ \ \ \ \ \ \ \ \ \ \ \ \ \ \ \ \ \ \ \ \ \ \ \ \ \ \ \ \ \ \ \ \ \ \ \ \ \ \ \ \ \ \ \ \ \ \ \ \ \ \ \ \ \ \ \ \ \ \ (3.8)

\bigskip The case where \ $g_{00}$\ depends also\ on \ $r$ $\ \ ,$
$\ \ \theta$\ \ and \ $\phi$\ \ \ was considered also by Berman (2008b)\ and
does not differ qualitatively from the present analysis, so that, we refer the
reader to that paper.

\bigskip

\bigskip{\LARGE 4. The exact solution to the Pioneers anomaly.}

Consider the possible solution for the rotating case.We equate (1.2) and
(3.8).We try a power-law solution for \ $R,$ \ \ and find,

\bigskip

$g_{00}=Ae^{t^{1-1/m}}$ \ \ \ \ \ \ \ \ \ \ \ \ \ \ \ \ \ ( $\ A=$constant \ \ ).

\bigskip

The scale-factor assumes a power-law , as in constant deceleration parameter
models (Berman,1983;--- and Gomide,1988),

\bigskip

$R=(mDt)^{1/m}$ \ \ \ \ \ \ \ \ \ \ \ \ \ \ \ \ \ \ \ \ \ \ \ \ \ \ \ \ \ \ \ \ \ \ \ \ \ \ \ \ \ \ ,\ \ \ \ \ \ \ \ \ \ \ \ \ \ \ \ \ \ \ \ \ \ \ \ \ \ \ \ \ \ \ \ \ \ \ \ \ (4.1)

\bigskip

where, $\ m$ $\ $, $\ \ D=$ \ constants, \ and,

\bigskip

$m=q+1>0$
\ \ \ \ \ \ \ \ \ \ \ \ \ \ \ \ \ \ \ \ \ \ \ \ \ \ \ \ \ \ \ \ \ \ \ \ \ \ \ \ \ \ \ \ \ ,
\ \ \ \ \ \ \ \ \ \ \ \ \ \ \ \ \ \ \ \ \ \ \ \ \ \ \ \ \ \ \ \ \ (4.2)

\bigskip

where \ $q$ \ \ \ is the deceleration parameter.\ \ \ We may choose
\ \ $q$\ \ \ \ as needed to fit the observational data.\ \ \ \ \ \ \ \ \ \ \ \ \ \ \ \ \ \ \ \ \ \ \ \ \ \ \ \ \ \ \ \ \ \ \ \ \ \ \ \ \ \ \ \ \ \ \ \ 

\bigskip We find,

$H=(mt)^{-1}$

\bigskip

If we now solve for energy-density of matter, and cosmic pressure, for a
perfect fluid, the best way to present the calculation, and the most simple,
is showing the matter energy-density $\ \ \ \rho$ \ \ \ \ and the
\ \ \ $\sigma$-or gravitational density parameter, to be defined below. We find,

$\rho=\frac{3t^{-2}}{m^{2}A\kappa}e^{-t^{1-1/m}}+3k(mDt)^{-2/m}-\frac{\Lambda
}{\kappa}$

\bigskip

$\sigma=\left(  \rho+3p-2\frac{\Lambda}{\kappa}\right)  =\left[
\frac{6(1-m)t^{-2}+3t^{-1/m}}{m^{2}A\kappa}\right]  e^{-t^{1-1/m}}$

\bigskip For the present Universe, the infinite time limit makes the above
densities become zero.

\bigskip It is possible to define,

\bigskip$\rho_{grav}=-\frac{3H^{2}}{\kappa g_{00}}$
\ \ \ \ \ \ \ \ \ \ \ \ \ \ \ \ \ \ \ \ \ \ \ \ \ \ (negative energy-density
of the gravitational field)

\bigskip Now, let us obtain the gravitational energy of the field,

\bigskip$E_{grav}=\rho_{grav}V=-(4/3)\pi R^{3}(\frac{3H^{2}}{\kappa g_{00}%
})=-\frac{c^{4}R^{4-3m}}{2Gm^{2}A(mD)^{1/m-3}}$ \ \ \ \ \ \ \ \ \ \ \ \ \ \ \ \ \ \ \ \ \ \ (4.3)

{\LARGE 5. The Second Pioneers Anomaly}

\bigskip

The universal angular acceleration, is given by,

\bigskip

$\alpha_{u}=\dot{\omega}=-cH/R=-c^{2}/R^{2}$ \ \ \ \ \ \ \ \ \ \ \ \ \ \ \ \ \ \ \ \ \ \ \ \ \ \ \ \ \ \ \ \ \ \ \ \ \ \ \ \ \ \ \ \ \ \ \ \ \ \ \ \ \ \ \ \ \ \ \ \ \ \ \ \ \ (5.1)

\bigskip\bigskip

\bigskip The spins of the Pioneers were telemetered, and as a surprise, it
shows that the on-board measurements yield a decreasing angular speed, when
the space-probes were not disturbed. Turyshev and Toth (2010), published the
graphs (Figures 2.16 and 2.17 in their paper), from which it is clear that
\ there is an angular deceleration of about 0.1 RPM per three years, or,

\bigskip

$\alpha\approx-1.2\times10^{-10}$rad/s$^{2}.$ \ \ \ \ \ \ \ \ \ \ \ \ \ \ \ \ \ \ \ \ \ \ \ \ \ \ \ \ \ \ \ \ \ \ \ \ \ \ \ \ \ \ \ \ \ \ \ \ \ \ \ \ \ \ \ \ \ \ (5.2)

\bigskip

As the diameter of the space-probes is about 10 meters, the linear
acceleration is practically the Pioneers anomalous deceleration value ,in this
case, -6.10$^{-8}$ cm.s$^{-2}{\LARGE .}$The present solution of the second
anomaly, confirms our first anomaly explanation.

\bigskip I have elsewhere pointed out that we are in face of an angular
acceleration frame-dragging field, for it is our result \ (5.1) above, for the
Universe, that causes the result (5.2), through the general formula,

$\alpha=-\frac{cH}{l}$ \ \ \ \ \ \ \ \ \ \ \ \ \ \ \ \ \ \ \ \ \ \ \ \ \ \ \ \ \ \ \ \ \ \ \ \ \ \ \ \ \ \ \ \ \ \ \ \ \ \ \ \ \ \ \ \ \ \ \ \ \ \ \ \ \ \ \ \ \ \ \ \ \ \ \ \ \ \ \ \ \ \ \ \ \ \ \ \ \ (5.3)

where \ $l$ \ \ \ is the linear magnitude of the localized body suffering the
angular acceleration frame-dragging.\bigskip\ For sub-atomic matter, this
angular acceleration can become important.

\bigskip{\LARGE 6. The Solution of the Fly-by Anomaly}

\bigskip

\bigskip Consider a two-body problem, relative to \ an inertial system. The
additional speed, measured at infinity, relative to the total speed, measured
at infinity, is proportional to twice the tangential speed of the earth,
$\ \ \ w_{e}R_{e}$, \ divided by the total speed $V\rightarrow wR\approx c$
\ \ taken care of the Universe angular speed. In fact, we write,

\bigskip

$\frac{\Delta V}{V}=\frac{V+\omega_{e}R_{e}-(V-\omega_{e}R_{e})}{c}%
=\frac{2\omega_{e}R_{e}}{c}\approx3\times10^{-6}$ \ \ \ \ \ \ \ \ \ \ \ \ \ \ \ \ \ \ \ \ \ \ \ \ \ \ \ \ \ (6.1)

\bigskip The trick, is that infinity, in a rotating Universe, like ours, has a
precise meaning, through the angular speed formula (1.2).

\bigskip

\bigskip{\LARGE 7. The Godlowski et al. Rotation}

\bigskip We, now, shall follow an idea by Godlowski et al.(2004), and supply
another General Relativistic model, of an expanding and rotating Universe.
Their idea, is that the homogeneous and isotropic models, may still rotate
relative to the local gyroscope, by means of a simple replacement, in the
Friedman-RW%
\'{}%
s equations, of the kinetic term, by the addition of a rotational kinetic one.

\bigskip

Einstein%
\'{}%
s field equations, for a perfect fluid with perfect gas equation of state, and
RW%
\'{}%
s metric, are two ones. The first, is an energy-density equation, the second
is a definition of cosmic pressure, which can be substituted by energy
momentum conservation. But, upon writing the \ \ \ $\dot{R}^{2}$ \ \ \ term,
we shall add an extra rotational term, namely $(\omega R)^{2}$, in order to
account for rotation. If we keep (3.1), the field equations become, for a flat Universe,

\bigskip

$6H^{2}=\kappa\rho+\Lambda$ \ \ \ \ \ \ \ \ \ \ \ \ \ \ \ \ \ \ \ \ \ \ \ \ \ \ \ \ \ \ \ \ \ \ \ \ \ \ \ \ \ \ \ \ \ \ \ \ \ \ \ \ \ \ \ \ \ \ \ \ \ \ \ \ \ \ \ \ \ \ \ \ \ \ \ \ \ \ \ \ \ \ \ \ \ \ \ \ \ \ \ (7.1)

\bigskip

with,

\bigskip$p=\beta\rho$ \ \ \ \ \ \ \ \ \ \ \ \ \ \ \ \ \ \ \ \ \ \ \ \ \ \ \ \ \ \ \ \ \ \ \ \ \ \ \ \ \ \ \ \ \ \ \ \ \ \ \ \ \ \ \ \ \ \ \ \ \ \ \ \ \ \ \ \ \ \ \ \ \ \ \ \ \ \ \ \ \ \ \ \ \ \ \ \ \ \ \ \ \ \ \ \ \ \ \ \ \ \ (7.2)

and,

\bigskip

$\dot{\rho}=-3\sqrt{2}H\rho(1+\beta)$ \ \ \ \ \ \ \ \ \ \ \ \ \ \ \ \ \ \ \ \ \ \ \ \ \ \ \ \ \ \ \ \ \ \ \ \ \ \ \ \ \ \ \ \ \ \ \ \ \ \ \ \ \ \ \ \ \ \ \ \ \ \ \ \ \ \ \ \ \ \ \ \ \ \ \ \ \ \ \ \ \ \ (7.3)

\bigskip

\bigskip The usual solution, with Berman%
\'{}%
s constant deceleration parameter models, render (Berman, 1983;--and Gomide, 1986),

\bigskip

$R=(mDt)^{1/m}$ \ \ \ \ \ \ \ \ \ \ \ \ \ \ \ \ \ \ \ \ \ \ \ \ \ \ \ \ \ \ \ \ \ \ \ \ \ \ \ \ \ \ \ \ \ \ \ \ \ \ \ \ \ \ \ \ \ \ \ \ \ \ \ \ \ \ \ \ \ \ \ \ \ \ \ \ \ \ \ \ \ \ \ \ \ \ \ \ \ (7.4)

\bigskip

$H=(mt)^{-1}$ \ \ \ \ \ \ \ \ \ \ \ \ \ \ \ \ \ \ \ \ \ \ \ \ \ \ \ \ \ \ \ \ \ \ \ \ \ \ \ \ \ \ \ \ \ \ \ \ \ \ \ \ \ \ \ \ \ \ \ \ \ \ \ \ \ \ \ \ \ \ \ \ \ \ \ \ \ \ \ \ \ \ \ \ \ \ \ \ \ \ \ \ \ (7.5)

\bigskip

$\ddot{R}=-qH^{2}R=-(m-1)H^{2}R$ \ \ \ \ \ \ \ \ \ \ \ \ \ \ \ \ \ \ \ \ \ \ \ \ \ \ \ \ \ \ \ \ \ \ \ \ \ \ \ \ \ \ \ \ \ \ \ \ \ \ \ \ \ \ \ \ \ \ \ \ \ \ \ \ \ \ \ (7.6)

\bigskip

Notice that we may have \ a negative deceleration parameter, implying that the
Universe accelerates, probably due to a positive cosmological "constant", but,
nevertheless, it is subjected to a negative rotational deceleration, a kind of
centripetal one, that acts on each observed point of the Universe, relative to
each observer, given by relation (1.2), so that,

\bigskip

\bigskip

$\ddot{R}=-qH^{2}R=qa_{cp}$ \ \ \ \ \ \ \ \ \ \ \ \ \ \ \ \ \ \ \ \ \ \ \ \ \ \ \ \ \ \ \ \ \ \ \ \ \ \ \ \ \ \ \ \ \ \ \ \ \ \ \ \ \ \ \ \ \ \ \ \ \ \ \ \ \ \ \ \ \ \ \ \ \ \ \ \ \ \ (7.7)

\bigskip

\bigskip We now supply the necessary relations among the constants, so that
the above equations be observed, namely,

$m=\frac{3}{2}\sqrt{2}(1+\beta)=\pm\frac{\sqrt{6}}{\sqrt{\kappa\rho
_{0}+\Lambda_{0}}}$

\bigskip

$\rho=\rho_{0}t^{-2}$ \ \ \ \ \ \ \ \ \ \ \ \ \ \ \ \ \ \ \ \ \ \ \ \ \ \ \ \ \ \ \ \ \ \ \ \ \ \ \ \ \ \ \ \ \ \ \ \ \ \ \ \ \ \ \ \ \ \ \ \ \ \ \ \ \ \ \ \ \ \ \ \ \ \ \ \ \ \ \ \ \ \ \ \ \ \ \ \ \ \ (7.8)

\bigskip

$\Lambda=\Lambda_{0}t^{-2}$ \ \ \ \ \ \ \ \ \ \ \ \ \ \ \ \ \ \ \ \ \ \ \ \ \ \ \ \ \ \ \ \ \ \ \ \ \ \ \ \ \ \ \ \ \ \ \ \ \ \ \ \ \ \ \ \ \ \ \ \ \ \ \ \ \ \ \ \ \ \ \ \ \ \ \ \ \ \ \ \ \ \ \ \ \ \ \ \ \ \ \ \ \ \ \ (7.9)

\bigskip

\bigskip

\bigskip{\LARGE 8. Final comments}

If we calculate the centripetal acceleration corresponding to the above
angular speed (1.2), we find, for the present Universe, with $R\approx10^{28}%
$cm and $\ c\simeq3.10^{10}$cm./s\ \ ,

\bigskip

$a_{cp}=-\omega^{2}R\cong-9.10^{-8}cm/s^{2}$
\ \ \ \ \ \ \ \ \ \ \ \ \ \ \ \ \ \ \ . \ \ \ \ \ \ \ \ \ \ \ \ \ \ \ \ \ \ \ \ \ \ \ \ \ \ \ \ \ \ \ \ \ \ \ \ \ \ \ \ \ \ \ \ \ \ \ \ (8.1)

\bigskip

Our model of Section 4 has been automatically calculated alike with (1.2) and
(8.1). This value matches the observed experimentally deceleration of the NASA
Pioneers' space-probes. Equation (3.3) shows that one can have a positive
cosmological lambda term accelerating the Universe, i.e., $\ \ \ddot{R}\geq0$
\ \ \ along with a centripetal deceleration that is felt by any observer,
relative any observed point, given by (8.1).Berman and Gomide (2010) had
obtained a Machian General Relativistic solution, though particular.We call it
Machian, because it parallels the semi-relativistic Machian solution by Berman (2007b).

\bigskip A cosmologist has made very important criticisms on our work. First,
he says why do not the planets in the solar system show the calculated
deceleration on the Pioneers? The reason is that elliptical orbits are closed,
and localized. You do not feel the expansion of the universe in the sizes of
the orbits either. In General Relativity books, authors make this explicit.
You do not include Hubble%
\'{}%
s expansion in Schwarzschild%
\'{}%
s metric. But, those space probes that undergo hyperbolic motion, which orbits
extend towards infinity, they acquire cosmological characteristics, like, the
given P.A. deceleration. Second objection, there are important papers (Rievers
and L\"{a}mmerzahl, (2011); Francisco et al. (2011); Cuesta (2011)) which
resolve the P.A. with non-gravitational Physics. Our answer, that is OK, we
have now alternative explanations. However, in the Introduction of this paper,
we have responded why thermal emission is no good an explanation because it
\ does not explain the other two anomalies neither why the elliptical orbiters
did not suffer the same deceleration; as to Cuesta (2011) he also has no
explanation for the other two anomalies. This does not preclude ours. Third,
cosmological reasons were discarded, including rotation of the Universe. The
problem is that those discarded cosmologies, did not employ the correct
metric. For instance, they discarded rotation by examining Godel model, which
is non expanding, and with a strange metric. The two kinds of rotating and
expanding metrics we employ now, were not discarded or discussed by the
authors cited by this cosmologist. Then, the final question, is how come that
a well respected author, dismissed planetary Coriolis forces induced by
rotation of distant masses, by means of the constraints in the solar system.
The answer is the same above, and also that one needs to consider Mach%
\'{}%
s Principle on one side, and the theoretical meaning of vorticities, because
one is not speaking in a center or an axis of rotation or so. When we say, in
Cosmology, that the Universe rotates, we mean that there is a field of
vorticities, just that. The whole idea is that Cosmology does not enter the
Solar System except for non-closed orbits that extend to outer space.\textit{
}For the Gomide Uehara RW%
\'{}%
s metric, it is the tri-space that rotates relative to the orthogonal time-axis.

\bigskip Another cosmologist pointed out a different "problem". He was
discussing the prior paper, to the present one (Berman and Gomide, 2010). He
objects, that the angular speed \ formula of ours, is coordinate dependent.
Now, when you choose a specific metric, you do it thinking about the kind of
problem you have to tackle. After you choose the convenient metric, you forget
tensor calculus, and you work with coordinate-dependent relations. They work
only for the given metric, of course.

\bigskip The solutions of Section 4, and Section 7, are in fact a large class
of solutions, for they embrace any possible deceleration parameter value, or,
any power-law \ scale-factor. Our solution with the rotation of the Universe,
is the only unified  explanation that applies to the three NASA anomalies.

{\LARGE \bigskip Acknowledgements}

\bigskip The authors thank Marcelo Fermann Guimar\~{a}es, Nelson Suga, Mauro
Tonasse, Antonio F. da F. Teixeira, and for the encouragement by Albert,
Paula, and Luiza Mitiko Gomide.

\bigskip{\Large References }

\bigskip Adler, R.J.; Bazin, M.; Schiffer, M. (1975) - \ \textit{Introduction
to General Relativity, }2$^{nd}$ Edition, McGraw-Hill, New York.

\bigskip\ Anderson,J.D. et al.(2002)-\textit{ "Study of the anomalous
acceleration of Pioneer 10 and 11".-} Phys. Rev. D 65, 082004 (2002

\bigskip Anderson,J.D. et al.(2008)-PRL100,091102.

Berman, M.S. (1983) - \textit{Special Law of Variation for Hubble%
\'{}%
s Parameter ,}Nuovo Cimento \textbf{74B, }182-186.

Berman, M.S. (2007) - \textit{Introduction to General Relativistic and
Scalar-Tensor Cosmologies, }Nova Science Publisher, New York. (see Section 7.12)

Berman, M.S. (2007b) - \textit{The Pioneer Anomaly and a Machian Universe} -
Astrophysics and Space Science, \textbf{312}, 275. Los Alamos Archives, http://arxiv.org/abs/physics/0606117.

Berman, M. S. (2008a) - \textit{A General Relativistic Rotating Evolutionary
Universe, }Astrophysics and Space Science, \textbf{314, }319-321.

Berman, M. S. (2008b) - \textit{A General Relativistic Rotating Evolutionary
Universe - Part II, }Astrophysics and Space Science, \textbf{315, }367-369.

Berman, M.S.; Gomide, F.M. (1988) - \textit{Cosmological Models with Constant
Deceleration Parameter - }GRG \ \textbf{20,}191-198.

\bigskip Berman, M.S.; Gomide, F.M. (2010) - Los Alamos Archives
arXiv:1011.4627v3. Submitted.

Berman,M.S.; Gomide, F.M.(2011) - \textit{On the Rotation of the Zero-Energy
Expanding Universe},Chapter 12, in "The Big-Bang:Theory, Assumptions and
Problems", ed. by O`Connell and Hale, Nova Science, N.Y., to be published.

Berman,M.S.; Gomide, F.M.(2011a) - \textit{Relativistic Cosmology and the
Pioneers Anomaly}, Los Alamos Archives, arXiv:1106.5388 .

Chechin,L.M. (2010)-Astron. Rep.54,719.

Godlowski, W.et al. (2004) - Los Alamos Archives, \ arxiv:astro-ph/0404329 .

Francisco, F. et al. (2011) - Los Alamos Archives, \ arXiv:1103.5222 .

Cuesta, H.J.M. (2011) - Los Alamos Archives, arXiv:1105.2832 .

Gomide, F.M.; Uehara, M. (1981) - Astronomy and Astrophysics, \textbf{95}, 362.

\bigskip L\"{a}mmerzahl, C. et al.(2008) - in \textit{Lasers,Clocks, and
dragg-free Control, }ed. by Dittus,Lammerzahl and Turyshev,Springer, Heidelberg.

MTW - Misner, C.W.; Thorne, K.S.; Wheeler, J.A. (1973)\ - \textit{Gravitation}%
, Freeman, San Franscisco.

\bigskip Ni,Wei-Tou(2008) - Progress Theor. Phys.Suppl.\textbf{172,}49-60.

Ni,Wei-Tou (2009) - Int.Journ.Modern Phys.\textbf{A24,}3493-3500.

Rievers, B.; Lammerzahl, C. (2011) - Los Alamos Archives, arXiv:1104.3985 .

Sabbata, de V.; Gasperini, M. (1979) - \textit{Lettere al Nuovo Cimento,
}\textbf{25}, 489.

So, L.L.; Vargas, T. (2006) - Los Alamos Archives, gr-qc/0611012 .

Su,S.C.;Chu,M.C.(2009)-Ap.J.703,354

Turyshev,S.G.;Toth,V.T. (2010) - \textit{The Pioneer Anomaly, }Los alamos
Archives, arXiv:1001.3686

Weinberg, S.\ (1972) - \textit{Gravitation and Cosmology, }Wiley. New
York.\bigskip

Wilson,T.L.;Blome,H.-J.(2008)-arXiv:0508.4067
\end{document}